
%
%
%
\input jnl
\input reforder
\input eqnorder

\baselineskip 18pt

\centerline{\bf{Comment on the Coupling of Zero Sound}}
\medskip
\centerline{\bf{to the ${\bf J=1^-}$ Modes of $^3$He-B}}
\vskip 1.0truein
\centerline{Ross H. McKenzie$^*$ and J. A. Sauls}
\medskip
\centerline{Department of Physics and Astronomy}
\centerline{Northwestern University}
\centerline{Evanston, IL \ 60208}
\vskip1.25truein
\centerline{\uppercase{\bf abstract}}
\medskip

Features in the zero sound attenuation near the pair-breaking edge
in superfluid $^3$He-B have been observed in large magnetic fields.
Schopohl and Tewordt [{\sl J. Low Temp. Phys.} {\bf 57}, 601 (1984)]
claim that the
$J=1^-, M=\pm 1$ order-parameter collective modes
couple to zero sound as a result of the distortion of the equilibrium
order parameter by a magnetic field; they identify the new features
with these modes. However, we show that, when the effect of gap
distortion on the collective modes is properly taken into account, the
collective mode equations of Schopohl and Tewordt yield no direct coupling of
zero sound to the $J=1^-$ modes. Thus, the identification of the
absorption features reported by Ling, Saunders and Dobbs
[{\sl Phys. Rev. Lett.} {\bf 59}, 461 (1987)]
near the pair-breaking edge with the $J=1^-$ modes is not
clearly established.
\vskip 1truein

{\narrower\smallskip\noindent * Present address:  Department of Physics,
Ohio State University, 174 West 18th Ave., Columbus, OH  43210, USA.}
\bigskip

\break

\baselineskip 21pt

Two features, a ``peak'' and an ``anti-peak'', have been
observed\refto {dan,lin,sau} in the attenuation spectrum of zero sound in
superfluid $^3$He-B near the pair-breaking edge ($\omega=2\Delta$)
in strong magnetic
fields. Schopohl and Tewordt\refto {ST84} (ST) have identified
these features with the
order-parameter collective modes that are odd under the particle-hole
transformation, have total angular momentum
$J=1$ and magnetic quantum numbers $M=\pm 1$.
In zero field elementary symmetry arguments show that the $J=1^-$ modes
do not couple to zero sound.\refto{ser83b}
To identify the observed features with the $J=1^-, M=\pm 1$
collective modes it must be shown that these modes have frequencies
corresponding to the observed positions in the spectrum,
and that they couple to zero sound in the presence of a magnetic field.

Schopohl and Tewordt show that in a magnetic field, $\vec H = H\hat z$,
the $J=1^-$ modes have frequencies
$$\omega _{1^-,M} = 2\Delta - M\ g\ \omega _L\ , \eqno(aa)$$
where $\omega _L$ is the effective Larmor frequency and $g$ is the
Land\'e g-factor for these modes.
The value for $g$ calculated by ST agrees with the value deduced
from the splitting of the two features seen by Ling, Saunders and
Dobbs.\refto{lin}
In addition, ST argue that the $J=1^-$, $M=\pm 1$ modes couple to zero
sound in a magnetic field. In a field the B-phase order parameter
is no longer the isotropic Balian-Werthamer (BW) state with $J=0$,
but has the form
$$\vec\Delta (\widehat p) = (\Delta_1\ \widehat p_x ,
\Delta_1\ \widehat p_y , \Delta_2\ \widehat p_z) \eqno(ab)$$
where the anisotropy (``gap distortion'') is quadratic in the field, {\it i.e.}
$\Delta_1 - \Delta_2  \sim H^2$. Schopohl and Tewordt claim that the
$J=1^-, M=\pm 1$ modes couple to zero sound in a magnetic field
because of this gap distortion. However,
ST neglected the distortion of the
collective modes by the field. When gap distortion is
properly included in the time-dependent gap equation the coupling
between zero sound and the $J=1^-$ modes considered by ST vanishes.
Thus, the identification of the attenuation features
observed by Ling, {\it et al.} with the $J=1^-$ modes is not
established. Further inconsistencies between the observations of
Ref. \cite{lin} and the theory of ST are pointed out in Ref. \cite{sau}.

The $J=1^-$ modes are excitations of the
the $\ell=1$ component of the imaginary part of the
order parameter, $d_\mu^- (\widehat p, \omega,\vec q \> ) =
d_{\mu j}^-(\omega,\vec q \> )\widehat p_j$, the dynamics of
which are described by the time-dependent gap equation\refto{rmck}
$$\int {d\Omega \over 4\pi}\ \lambda (\omega,\eta)
\biggl[(\omega ^2-\eta ^2-4\Delta ^2)d^-_\mu + 4(\vec
\Delta \cdot \vec d^-)\Delta_\mu + G\omega_L (\hat z
\cdot \vec \Delta) (\vec \Delta \times \vec d^-)_\mu \biggr]
\widehat p_j$$ $$ = \int {d\Omega \over 4\pi}
\lambda(\omega,\eta)\Delta_\mu \widehat p_j \biggl[\omega
\varepsilon^+ + \eta \varepsilon^-\biggr] \eqno(ac)$$
where $\eta = v_f\ \widehat p\cdot \vec q,\ \lambda(\omega, \eta)$ is
the Tsuneto function, $G$ is a function which determines the Land\'e
factors of the modes, and $\varepsilon^+$ and $\varepsilon^-$ are mean
fields related to density and current oscillations, respectively.
The left side of \(ac) determines the frequencies of the order
parameter collective modes and the right side determines how these
modes couple to the density and current oscillations induced by zero
sound.\refto{note}  Equation \(ac) is equivalent to eq. (15) of ST in
Ref. \cite{ST81} and is the starting point for our discussion of ST's
calculation
in Ref. \cite{ST84}.

First consider the zero-field case. The order parameter is
the isotropic BW state, $\vec \Delta (\widehat p) =
\Delta \ \widehat p$, and eq. \(ac) is solved by decomposing
$d^-_{\mu j}$ in terms of the spherical tensors, $t^{J,M}_{\mu j}$,
defined by eqs. (108) and (109) in Ref. \cite{rmck},
$$d^-_{\mu j}(\omega, \vec q \> ) = \sum_{J=0}^2 \sum_{M=-J}^{+J}
D_{J,M}(\omega, \vec q \> ) \> t^{J,M}_{\mu j}\ . \eqno(ad) $$
If we neglect the dispersion of the modes, then eq. \(ac)
decomposes into nine independent equations with left side of the form
$$(\omega^2 - \omega^2_{J,M})\ D_{J,M}(\omega,0)\ .$$
This decomposition is equivalent to ST's diagonalization of their matrix
$R$ by means of a unitary transformation. Thus, $J$ and $M$ are good
quantum numbers for the collective modes of the BW state at zero
wavevector. However, when the gap $\vec \Delta$ is distorted in
a strong magnetic field, the equilibrium order parameter, when
represented as a tensor, $\Delta_{\mu} = \Delta_{\mu i}\ \widehat p_i$,
is composed of $J=0$ {\it and} $J=2$ spherical tensors.
Similarly, $J$ is no longer a good quantum number for the collective
modes.  This is an important point because ST neglected this change in
symmetry of the collective modes in their calculation of the coupling to
zero sound; they assumed that in  a strong magnetic field the modes
near $2\Delta$ are
described by the same $J=1$ tensor as in zero field.

To examine the solutions of eq. \(ac) when the gap is distorted, we
expand $d^-_{\mu}(\widehat p)$ as
$$d^-_\mu(\widehat p,\omega, \vec q \> ) =
{\cal D}_{\mu j}(\omega, \vec q \> ) \Delta_j(\widehat p)\ , \eqno(ae)$$
and then separate ${\cal D}_{\mu j}$ into symmetric, $S_{\mu j}$, and
antisymmetric, $A_{\mu j}$, parts,
$${\cal D}_{\mu j} = S_{\mu j} + A_{\mu j}. \eqno(af)$$
Furthermore, we can write $A_{\mu j} = \sum_{M=-1}^{+1} A^M
t^{1,M}_{\mu j}$, where $t^{1,M}_{\mu j} ={1 \over \sqrt{2}} \epsilon_{\mu j k}
\ u^M_k$ with $u^0_k = \widehat z_k$ and $u^{\pm 1}_k ={1 \over
\sqrt{2}}(\widehat x \pm i \widehat y)_k$,  so that
$$d^-_\mu = \sum _{M=-1}^{M=+1} A^M (\vec u^M \times \vec \Delta)_\mu +
S_{\mu j} \Delta_j \ .\eqno(ag)$$
To obtain the equation describing the dynamics of the antisymmetric
part, $A_{\mu j}$, we contract \(ac) with $t_{\mu k}^{1,-M} \Delta_{kj}$.
This gives our main result:
$$\int {d\Omega \over 4\pi}\ \lambda (\omega,\eta)
\Big\{\omega ^2-2g\omega \omega_L M-\eta ^2-4\Delta ^2\Big\} A^M
|\vec u^M \times \vec \Delta |^2 \ + \ F^M$$
$$ = \int {d\Omega \over 4\pi}
\lambda(\omega,\eta)\{(\vec u^M \times \vec \Delta) \cdot
\vec\Delta\}\ (\omega\varepsilon^++\eta\varepsilon^-)\equiv 0\ ,\eqno(ah)$$
where $F^M$ describes mixing with the $J=2^-,M\pm 1$ modes.
The right side of eq. \(ah) shows that the
anti-symmetric parts, which have frequencies near 2$\Delta$, are
{\it not} driven by longitudinal zero sound.
Schopohl and Tewordt obtained a nonzero coupling because
they incorrectly assumed that the
order-parameter oscillations were parallel to $\vec u^M \times \widehat p$ ;
thus, neglecting the field distortion of the collective modes. On
the right side of eq. \(ah) ST have a term proportional to
$\vec \Delta \cdot (\widehat p \times \widehat u^M)$, which is
proportional to
$\Delta _1 - \Delta _2 \propto H^2$, for $M = \pm1$.
However, the driving term is precisely zero
because the order-parameter oscillations
are parallel to $\vec u^M \times \vec \Delta$ and thus
orthogonal to the equilibrium order parameter, $\vec
\Delta$, even when it is distorted by strong magnetic fields.

There is an indirect coupling of sound to the $J=1^-$ modes, which was
not considered by ST, that arises from the off-resonant excitation of the
$J=2^-$, $M=\pm 1$ modes represented by $F^M$.
However, this term gives a much smaller coupling, and a field dependence
for the coupling that is in disagreement with experiment.
To leading order in $q^2$,
$$F^{\pm 1} ={ i \over 6} \lambda (\omega)
(\omega ^2-4 \Delta ^2) (\Delta _2^2-\Delta _1^2) D_{2,\pm1}
(\omega, \vec q \> )\ . \eqno(aj) $$
The $J=2^-,M \pm 1$ modes couple to longitudinal sound for
fields at an oblique angle relative to the sound propagation direction.
However, because of the prefactor $(\omega ^2-4 \Delta ^2)$ in $F^{\pm 1}$
this indirect coupling of sound to the $J=1^-$ modes is of order $H^3$.

To summarize, we have shown that the collective modes of the imaginary part of
the
order parameter with frequency near $2\Delta$ do not couple to
zero sound by the mechanism proposed by Schopohl and Tewordt.  However,
there may be other mechanisms
which couple these modes to longitudinal sound, {\it e.g.} the indirect
coupling via the $J=2^-$, $M=\pm 1$ modes. Thus,
the $J=1^-$ modes may still be identifiable
with the two features in the sound attenuation observed near
the pair-breaking edge\refto {dan,lin,sau}; however, the theory of the
coupling mechanism remains to be worked out.

\bigskip
\centerline{ACKNOWLEDGEMENTS}

We thank Bill Halperin and Geneva Moores for their comments on this
paper. This work was supported in part by the National Science Foundation
through the
Northwestern University Materials Science Center, Grant No. DMR8821571,
and in part by the DOE - Basic Energy Sciences, Division of Material Sciences.

\vfill
\eject

\references

\refis{ser83b}J. W. Serene, in {\it Quantum Fluids and Solids -1983}
(American Institute of Physics, New York, 1983), edited by
E. D. Adams and G. G. Ihas, p. 305.

\refis{lin}R. Ling, J. Saunders, and E. R. Dobbs, {\sl Phys. Rev. Lett.}
{\bf 59}, 461 (1987).

\refis{ST84}N. Schopohl and L. Tewordt, {\sl J. Low Temp. Phys.}
{\bf 57}, 601 (1984).

\refis{ST81}N. Schopohl and L. Tewordt, {\sl J. Low Temp. Phys.}
{\bf 45}, 67 (1981).

\refis{note}We note that for
$q \not= 0$ the collective modes which couple to zero sound are driven not
just by the density and current oscillations on the right side of
\(ac), but also by the oscillations in the phase of the order parameter
(the $J=0^-$ mode). ST neglected the latter in
both Refs. \cite{ST84} and \cite{ST81}
because they incorrectly assumed that for
$q \not= 0$ modes with different total angular momentum do not couple.
Consequently, the coupling constants which they obtained disagree with
the results of other authors. For example, in Ref. \cite{ST81},
ST obtain a coupling constant for the $J=2^-$ modes, to
first order in $q^2$, that is proportional to
${\partial \lambda (\omega,\eta) \over \partial \eta ^2 }\bigg|_{\eta=0},$
whereas W\"olfle [{\sl Phys. Rev. B} {\bf 14}, 89 (1976)]
and Maki [{\sl J. Low Temp. Phys.} {\bf 16}, 465 (1974)]
independently find the coupling to be proportional to
$\lambda (\omega,0)$ because they include the coupling to phase oscillations.

\refis{sau}J. Saunders, R. Ling, W. Wojtanowski, and E. R. Dobbs,
{\sl J. Low Temp. Phys.} {\bf 79}, 75 (1990).

\refis{dan}M. E. Daniels, E. R. Dobbs, J. Saunders, and P. L. Ward,
{\sl Phys.  Rev. B} {\bf 27}, 6988 (1983).

\refis{rmck}See R. H. McKenzie and J. A. Sauls, in {\sl Helium Three}
(Elsevier, Amsterdam, 1990), edited by W.P. Halperin and L.P.  Pitaevskii, for
the notation and a derivation of eq. \(ac).

\endreferences
\end